\documentclass[prb, 11pt, superscriptaddress, notitlepage, nofootinbib]{revtex4-1}
\usepackage{natbib}
\usepackage{graphicx}
\usepackage{dcolumn}
\usepackage{amsmath}
\usepackage{amssymb}
\usepackage{subfigure}
\usepackage[dvipsnames]{xcolor}
\usepackage{bm}
\usepackage{siunitx}

\usepackage{tikz}
\usetikzlibrary{arrows,positioning, decorations.markings} 
\tikzstyle{FullG} = [thick, decoration={markings,mark=at position
   0.58 with {\arrow[ultra thick]{stealth'}}},
   double distance=0.8pt,
   preaction = {draw},
   postaction = {decorate}]
\tikzset{
    >=stealth',
    punkt/.style={
           rectangle,
           rounded corners,
           draw=black, thick,
           minimum height=2em,
           text centered},
    punktd/.style={
           rectangle,
           rounded corners,
           draw=black, thick,
           minimum height=3.6em,
           align = center,
           text centered},
    pil/.style={
           ->,
           thick,
           shorten <=2pt,
           shorten >=2pt,}
}

\newcommand{\bff}{\mathbf{f}}
\newcommand{\bp}{\mathbf{p}}
\newcommand{\br}{\mathbf{r}}
\newcommand{\bn}{\mathbf{n}}

\newcommand{\bs}{\mathbf{s}}
\newcommand{\bj}{\mathbf{j}}

\newcommand{\bz}{\mathbf{e}_z}
\newcommand{\by}{\mathbf{e}_y}
\newcommand{\bx}{\mathbf{e}_x}

\newcommand{\bsigma}{\boldsymbol{\sigma}}
\newcommand{\bcalA}{\boldsymbol{\mathcal{A}}}
\newcommand{\bcalF}{\boldsymbol{\mathcal{F}}}
\newcommand{\bcalB}{\boldsymbol{\mathcal{B}}}
\newcommand{\taus}{\tau_s}
\newcommand{\tauDP}{\tau_\textnormal{\tiny DP}}

\newcommand{\lDP}{l_\textnormal{\tiny DP}}

\newcommand{\sigmaD}{\sigma_\mathrm{D}}

\newcommand{\qsg}{q_\mathrm{sg}}
\newcommand{\qisg}{q_\mathrm{isg}}

\newcommand{\eF}{\epsilon_F}
\newcommand{\pF}{p_F}

\newcommand{\gr}{{g_r^{\uparrow\downarrow}}}
\newcommand{\gi}{{g_i^{\uparrow\downarrow}}}

\newcommand{\qone}{q_1}
\newcommand{\qtwo}{q_2}
\newcommand{\tqone}{\tilde{q}_1}
\newcommand{\tqtwo}{\tilde{q}_2}

\newcommand{\mparallel}{m_\parallel}
\newcommand{\mperp}{m_\perp}
\newcommand{\Dparallel}{D_\parallel}
\newcommand{\Dperp}{D_\perp}
\newcommand{\tR}{d}

\begin{document}
\title{Quasiclassical theory of the spin-orbit magnetoresistance of three-dimensional Rashba metals}
\author{Sebastian T\"{o}lle}
\affiliation{Universit\"at Augsburg, Institut f\"ur Physik, 86135 Augsburg, Germany}
\author{Michael Dzierzawa}
\affiliation{Universit\"at Augsburg, Institut f\"ur Physik, 86135 Augsburg, Germany}
\author{Ulrich Eckern}
\affiliation{Universit\"at Augsburg, Institut f\"ur Physik, 86135 Augsburg, Germany}
\author{Cosimo Gorini}
\affiliation{Universit\"at Regensburg, Institut f\"ur Theoretische Physik, 93040 Regensburg, Germany}


\begin{abstract}
The magnetoresistance of a three-dimensional Rashba material placed on top of a ferromagnetic insulator is theoretically investigated. In addition to the intrinsic Rashba spin-orbit interaction, we also consider extrinsic spin-orbit coupling via side-jump and skew scattering, and the Elliott-Yafet spin relaxation mechanism. The latter is anisotropic due to the mass anisotropy which reflects the noncentrosymmetric crystal structure of three-dimensional Rashba metals. A quasiclassical approach is employed to derive a set of coupled spin-diffusion equations, which are supplemented by boundary conditions that account for the spin-transfer torque at the interface of the bilayer.
The magnetoresistance is fully determined by the current-induced spin polarization, i.e., it cannot in general be ascribed to a single (bulk) spin Hall angle. Our theoretical results reproduce several features of the experiments, at least qualitatively, and contain established phenomenological results in the relevant limiting cases. In particular, the anisotropy of the Elliott-Yafet spin relaxation mechanism plays a major role for the interpretation of the observed magnetoresistance.
\end{abstract}
\maketitle

\section{Introduction}
The fundamental tasks in the field of spintronics \cite{zutic2004,sinova2015} are to generate, manipulate, and detect spin densities or spin currents. 
One particularly interesting example where all these tasks are achieved simultaneously is the spin Hall magnetoresistance in a normal-metal/ferromagnet bilayer structure.\cite{nakayama2013, avci2015, han2014} 
In this case, an electric current in the normal metal generates a spin current via the spin Hall effect.\cite{dyakonov1971,hirsch1999,kato2004,wunderlich2005} This spin current gets only partly reflected at the interface to the adjacent ferromagnet, thereby exerting a torque on the magnetization.\cite{slonczewski1996, berger1996, tsoi1998} The reflected spin current is converted back into a charge current due to the inverse spin Hall effect,\cite{zhao2006,saitoh2006,valenzuela2006} resulting in a magnetization-dependent spin-orbit signature in the magnetoresistance.

Recently, a new type of spin-orbit-dependent magnetoresistance has gained considerable attention, the Rashba-Edelstein magnetoresistance.\cite{nakayama2016,nakayama2017} It relies on the inverse spin galvanic effect,\cite{ivchenko1978,vasko1979,aronov1989} a current-induced spin polarization due to spin-orbit coupling,\cite{kato2004_2,silov2004} also known as Edelstein or Rashba-Edelstein effect\cite{edelstein1990} in systems with Rashba spin-orbit coupling.\cite{rashba1960,bychkov1984} A typical experimental setup consists of a substrate/normal-metal/ferromagnet trilayer with a two-dimensional electron gas (2DEG) at the substrate/normal-metal interface. The magnetoresistance is usually explained as follows:\cite{nakayama2016} A current-induced spin polarization in the 2DEG leads to a spin current which flows through the normal metal, gets reflected at the normal-metal/ferromagnet interface, and is then converted back again to a charge current in the 2DEG via the spin galvanic effect.\cite{ganichev2001, ganichev2002} 

However, in dirty Rashba systems the interplay between extrinsic effects (due to impurities) and intrinsic effects (due to the band or device structure) leads to a non-trivial interaction of spin densities and spin currents.\cite{sinova2015, raimondi2012} Accordingly, the various spin-orbit signatures, e.g., via the spin galvanic and the (in-plane) inverse spin Hall effect,  in charge signals are hard to separate,\cite{grigoryan2014, toelle2017} eventually leading to a non-trivial magnetization dependence of the magnetoresistance.\cite{toelle2018} Additional contributions such as the anisotropic magnetoresistance in ferromagnetic metals, or spin Hall effects and/or a field-dependent magnetoresistance in the substrate/normal-metal part of the trilayer structure,\cite{nakayama2017} complicate the separation of Rashba-related effects from confounding signals.

One possibility to overcome these problems is to consider a bilayer consisting of a three-dimensional (3D) system with Rashba spin-orbit coupling and an insulating ferromagnet. Although commonly associated with (quasi) two-dimensional asymmetric systems, there exists a new class of bulk 3D Rashba metals\cite{ishizaka2011,niesner2016,martin2017} with rather strong Rashba spin-orbit coupling due to their noncentrosymmetric crystal structure. Obviously, these materials offer an interesting playground for investigations of Rashba-associated signatures in the charge sector, e.g., the anisotropy of the dc conductivity.\cite{brosco2017}
 
In this article, we theoretically investigate the magnetoresistance of such 3D Rashba metals, taking into account a mass anisotropy and both Dyakonov-Perel and Elliott-Yafet spin relaxation. To be consistent, we additionally consider extrinsic spin-orbit coupling via side-jump and skew scattering, hence our theory goes substantially beyond phenomenological approaches to the spin Hall magnetoresistance in heavy-metal/ferromagnet bilayers,\cite{chen2013} but recovers their results in the appropriate limiting cases.  Essentially, we show that in composite systems made of a ferromagnet and an anisotropic metal where Rashba and extrinsic spin-orbit coupling coexist, magnetoresistance signals are determined by current-induced spin polarizations.  In other words, such signals do not allow access to a single, well-defined (bulk) spin Hall angle, unless specific limiting conditions are met.

Our paper is organized as follows. We specify the boundary conditions and introduce the model of the system under consideration in Sec.\ \ref{Sec_model}. Section \ref{Sec_Spin} focuses on the current-induced spin polarization, paving the way for a consistent description of the magnetoresistance as presented in Sec.\ \ref{Sec_Magnetoresistance}. We briefly conclude in Sec.\ \ref{Sec_conclusion}.

\section{The model}\label{Sec_model}

\begin{figure}[tb]
\centering
\includegraphics[width=0.6\columnwidth]{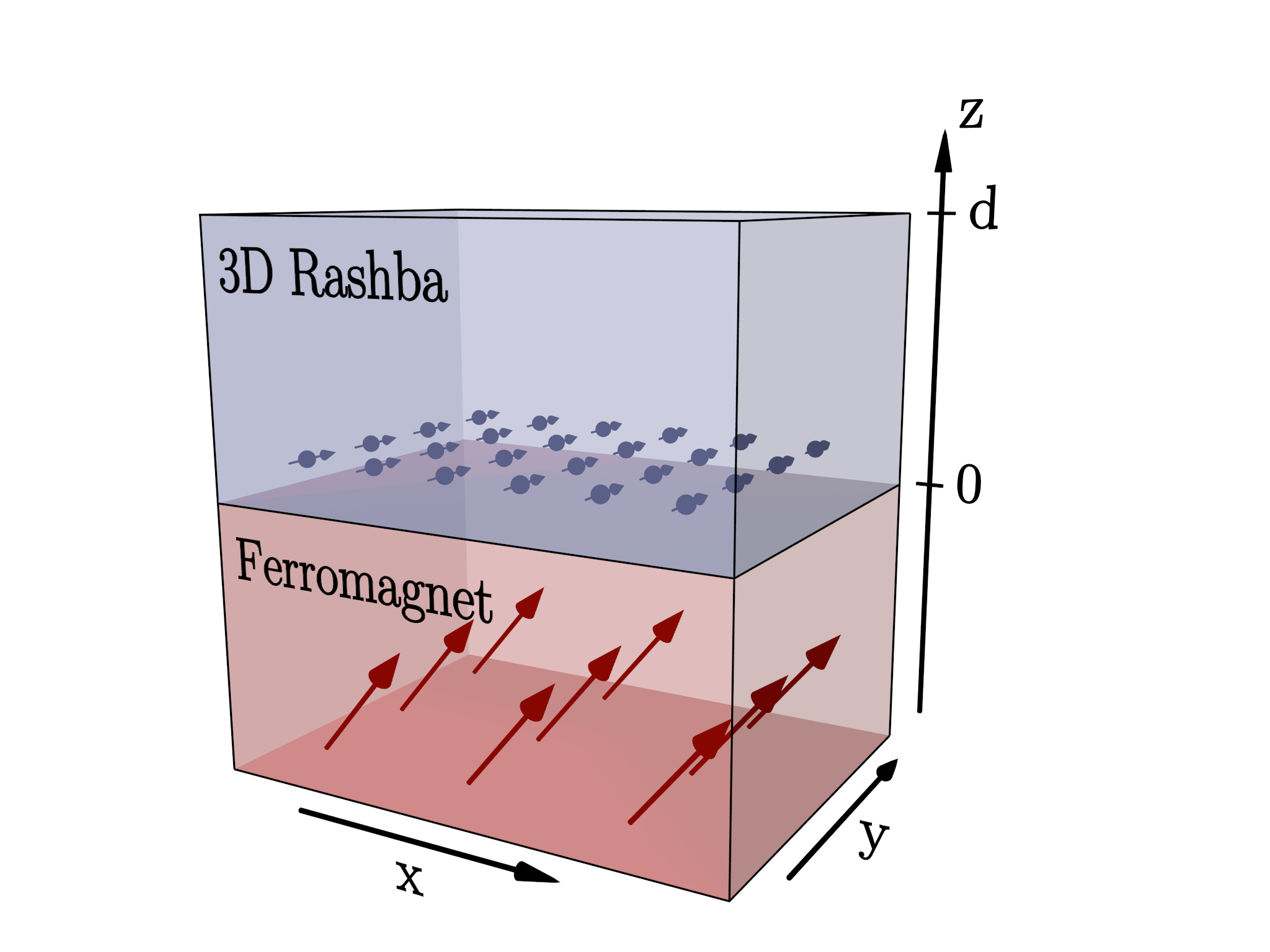}
\caption{Left: Schematic picture of the setup with a Rashba system (``3D Rashba'') placed on top of a ferromagnetic insulator (``Ferromagnet''). Right: Definition of the angles $\alpha$, $\beta$, and $\gamma$ which describe the rotation of the magnetization direction in the $xy$, $yz$, and $xz$ planes, respectively. \label{Fig_system}}
\end{figure}

The setup under consideration is schematically depicted in Fig.~\ref{Fig_system}. It consists of a 3D Rashba metal of thickness $\tR$ which is placed on top of a ferromagnetic insulator with the interface at $z=0$. The ferromagnet offers the possibility to manipulate the spin current across the interface due to the spin transfer torque by varying the magnetization direction $\bn$. The boundary condition for the spin current in the Rashba metal at the interface is given by\cite{kovalev2002, slonczewski2002}
\begin{equation}
\bj_z (0) = \frac{1}{2\pi \hbar N_0} \left[ \gr \bn \times \big( \bn \times \bs(0) \big) + \gi \bn \times \bs(0) \right] \, , \label{Eq_STT}
\end{equation}
where $\gr$ and $\gi$ are the real and the imaginary part of the spin mixing conductance. Here, $N_0 = \mparallel\sqrt{2\mperp\eF}/2\pi^2\hbar^3$ is the density of states per spin and volume at the Fermi energy of the 3D Rashba metal, which is described by the model Hamiltonian
\begin{equation} \label{Eq_Hamiltonian}
H = \frac{p_x^2 + p_y^2}{2\mparallel} + \frac{p_z^2}{2\mperp} - \frac{\alpha_R}{\hbar}  \left(\bp \times \bsigma  \right)  \cdot \bz  + H_\mathrm{imp} \, ,
\end{equation}
where $\alpha_R$ is the Rashba coefficient, and $\bsigma = (\sigma^x, \sigma^y, \sigma^z)$ is the vector of Pauli matrices. The inversion symmetry breaking direction $\bz$ accounts for the noncentrosymmetric crystal structure. Correspondingly, $\mparallel$ and $\mperp$ are the in-plane and out-of-plane effective masses. 
Disorder due to nonmagnetic impurities is taken into account by
\begin{equation}
H_\mathrm{imp} = V + \frac{\lambda^2}{4\hbar} \bsigma \cdot \left(\nabla V \times \bp \right) \, ,
\end{equation}
where $\lambda$ is the effective Compton wavelength, and $V$ is a $\delta$-correlated random potential.

Based on this microscopic model, we use a generalized Boltzmann equation for the distribution function $f=f^0 + \bff \cdot \bsigma$ as derived in Ref.~\onlinecite{gorini2010}. Here, $f^0$ and $\bff$ are the charge and the spin distribution functions which yield the spin density
\begin{equation}
\bs = \int \frac{d^3 p}{(2\pi\hbar)^3} \bff \, ,
\end{equation}
and the charge and spin current in $i = x,y,z$ direction,
\begin{align}
j_i &= -2e \int \frac{d^3p}{(2\pi\hbar)^3} v_i f^0 \, ,\\ 
\bj_i &=  \int \frac{d^3 p}{(2\pi\hbar)^3} v_i \bff \, , \label{Eq_Def_spincurrent}
\end{align}
where $v_i$ is the $i$-th component of the velocity $\mathbf{v} = (p_x/\mparallel, p_y/ \mparallel, p_z/\mperp)$. For technical details regarding the Boltzmann equation and the derivation of the transport equations used in the following sections we refer to App.~\ref{App_Boltzmann}.

Disorder, as taken into account by $H_\mathrm{imp}$, leads to momentum relaxation, $1/\tau$, and two types of spin relaxation: Dyakonov-Perel relaxation due to Rashba spin-orbit coupling, and Elliott-Yafet relaxation due to spin-orbit interaction with the random potential. Both relaxation mechanisms are anisotropic,
\begin{equation}
\partial_t \bs \sim - \frac{1}{\tauDP} \begin{pmatrix}
1 & 0 & 0 \\
0 & 1 & 0 \\
0 & 0 & 2 
\end{pmatrix}  \bs 
- \frac{1}{\taus} \begin{pmatrix}
1 & 0 & 0 \\
0 & 1 & 0 \\
0 & 0 & \zeta 
\end{pmatrix}  \bs \, ,
\end{equation}
where $1/\tauDP$ and $1/\taus$ are the Dyakonov-Perel and Elliott-Yafet relaxation rates, respectively. In the dirty regime, the former is given by
\begin{equation}
\frac{1}{\tauDP} = \left( \frac{2\mparallel\alpha_R}{\hbar^2} \right)^2\Dparallel
\end{equation}
with  the in-plane diffusion constant $\Dparallel = 2\eF\tau/3\mparallel$. The anisotropy of the Elliott-Yafet mechanism depends on the masses via the parameter
\begin{equation}
\zeta = \frac{2\mperp}{\mperp + \mparallel} \, ,
\end{equation} 
and the corresponding relaxation rate is given by
\begin{equation}
\frac{1}{\taus} = \frac{8}{9(2-\zeta)} \left( \frac{\lambda \pF}{2\hbar} \right)^4 \frac{1}{\tau}  \, ,
\end{equation}
with $\pF = \sqrt{2 \mparallel \eF}$.\footnote{A brief outline of the Elliott-Yafet spin relaxation as described within the Boltzmann theory is given in App.~\ref{App_Boltzmann}.} Note that $\tauDP/\taus \sim \tau^{-2}$ can be enhanced by increasing the temperature, since $\tau^{-1}$ is typically an increasing function of the temperature in a metallic system.

\section{Current-induced spin polarization}\label{Sec_Spin}
In this section, we investigate the current-induced spin polarization in the spin diffusive limit, in the sense that $\pF\tau / \mparallel \lDP \ll 1$, where $\lDP = \sqrt{\Dparallel \tauDP}$.
Neglecting spin-dependent contributions to the charge current (thus $j_x \approx \sigmaD E_x$, where $\sigmaD$ is the Drude conductivity)  the Boltzmann equation yields the following set of diffusion equations for the spin density:
\begin{align}
\qone^{2} s^x &=  \nabla_z^2 s^x \, , \label{Eq_sx_Diff} \\
\qone^2 s^y &=  \nabla_z^2 s^y + \qone^2 s_0^y \, , \label{Eq_sy_Diff} \\
\qtwo^2 s^z &=  \nabla_z^2 s^z \, . \label{Eq_sz_Diff}
\end{align}
The inverse spin relaxation lengths $\qone$ and $\qtwo$ are given by
\begin{align}
\qone &= \frac{1}{\lDP} \sqrt{ \frac{\mperp}{\mparallel}\left(1 + \frac{\tauDP}{\taus}\right)}   \, ,  \label{Eq_qperp} \\
\qtwo &= \frac{1}{\lDP} \sqrt{ \frac{\mperp}{\mparallel} \left(2 + \zeta\frac{\tauDP}{\taus}\right)}  \, .  \label{Eq_qparallel}
\end{align}
We have also introduced the bulk current-induced spin polarization in the homogeneous case, 
\begin{equation}
s_0^y = - \frac{1}{\Dparallel \qisg} \frac{\sigmaD E_x}{2e} \, ,	\label{Eq_iSGE}
\end{equation}
where $\qisg$ (``isg'' $=$ inverse spin galvanic) is defined by
\begin{equation}
\frac{1}{\Dparallel \qisg} = \frac{\taus/\lDP}{1+\taus/\tauDP} \left( \xi_\mathrm{int} \theta_\mathrm{int}^\mathrm{sH} + \theta_\mathrm{ext}^\mathrm{sH} \right) \, .
\end{equation}
Here, $\theta_\mathrm{int}^\mathrm{sH} = \alpha_R\tau/\hbar \lDP$ accounts for the Rashba contribution to the spin Hall angle,\footnote{Note that $\theta_\mathrm{int}^\mathrm{sH}$ is not the bulk spin Hall angle in case of a pure Rashba system.\cite{raimondi2012}} and
\begin{equation}
\xi_\mathrm{int} = 1- \frac{\tauDP}{\taus}\left(1-\frac{3\zeta}{4}\right) \, . \label{Eq_xi}
\end{equation}
Equation (\ref{Eq_iSGE}) describes the inverse spin galvanic effect in an anisotropic Rashba metal, explicitly taking into account side-jump and skew scattering via the parameter $\theta_\mathrm{ext}^\mathrm{sH}$, the extrinsic contribution to the spin Hall angle.
\footnote{More precisely, $\theta_\mathrm{ext}^\mathrm{sH} = \theta_\mathrm{sj}^\mathrm{sH} + \theta_\mathrm{ss}^\mathrm{sH} $, where $ \theta_\mathrm{sj,ss}^\mathrm{sH} = 2e\sigma_\mathrm{sj,ss}^\mathrm{sH}/\sigmaD$ with the side-jump and skew scattering contributions to the spin Hall conductivity, $\sigma_\mathrm{sj}^\mathrm{sH}$ and $\sigma_\mathrm{ss}^\mathrm{sH}$, being defined in Ref.~\onlinecite{raimondi2012}, respectively.}

To proceed, we explicitly solve Eqs.~(\ref{Eq_sx_Diff})--(\ref{Eq_sz_Diff}) by taking into account proper boundary conditions at $z=0$ and $z=\tR$. These are given by Eq.~(\ref{Eq_STT}) and the condition $\bj_z(\tR) = 0$, corresponding to spin-conserving scattering.
We obtain
\begin{equation} \label{Eq_sy_total}
s^y(z)= s_0^y + \Delta s_\mathrm{sc}^y(z) + \Delta s^y(z, \bn) \, ,
\end{equation}
where 
\begin{equation} \label{Eq_ssc}
\Delta s_\mathrm{sc}^y(z) = \frac{\theta_\mathrm{ext}^\mathrm{sH}}{\Dparallel \qone} \frac{ \sigmaD E_x}{2e } \frac{\sinh\boldsymbol{(} \qone(\tR/2 - z) \boldsymbol{)}}{\cosh(\qone\tR/2)} 
\end{equation}
is the spin accumulation which arises due to the spin current $j_z^y$ even in the absence of the ferromagnet. The magnetization-dependent contribution is given by $\Delta s^y(z, \bn)$. In the following, we focus on $\Delta s^y(z, \bn)$ with the magnetization vector lying in the $xy$, $yz$, or $xz$ plane, respectively, i.e., the $\alpha$, $\beta$, and $\gamma$ scans as defined in Fig.~\ref{Fig_system}.
After some algebra, we obtain
\begin{equation} \label{Eq_sy_abc}
\Delta s^y_{\alpha, \beta, \gamma}(z) = - A(z) f_{\alpha, \beta, \gamma} \, ,
\end{equation}
with 
\begin{equation}
 A(z) = s_0^y  \left[ 1-  \tanh\left( \qone \tR/2 \right) \theta_\mathrm{ext}^\mathrm{sH}  \frac{\qisg}{\qone} \right] \frac{\cosh\boldsymbol{(}\qone(d-z)\boldsymbol{)}}{\cosh(\qone d)} \, .
\end{equation}
The angular dependence is given by
\begin{widetext}
\begin{align} 
f_\alpha &= \frac{\left(q_i^2 + q_r^2 + q_r \tqtwo \right) \cos^2 \alpha }{q_i^2 + (\tqone + q_r)(\tqtwo + q_r)} \, , \label{Eq_f_alpha} \\
f_\beta &=  \frac{\left(q_i^2 + q_r^2 + q_r \tqone \right) \cos^2 \beta}{ \frac{\tqone}{\tqtwo} \left[ q_i^2 + (\tqone+q_r)(\tqtwo+q_r) \right] \sin^2 \beta  + \left[q_i^2 +(\tqone + q_r)^2 \right] \cos^2\beta }  \, ,  \label{Eq_f_beta} \\
f_\gamma &= \frac{q_i^2 + q_r^2 + q_r \tqone - \left(q_i^2 + q_r^2\right) \left(1- \frac{\tqone}{\tqtwo}\right) \sin^2 \gamma  }{ \frac{\tqone}{\tqtwo} \left[ q_i^2 + (\tqone+q_r)(\tqtwo+q_r) \right] \sin^2 \gamma  + \left[q_i^2 +(\tqone + q_r)^2 \right] \cos^2\gamma } \, , \label{Eq_f_gamma}
\end{align}
\end{widetext}
where the respective scan is indicated by the subscript.
Furthermore, we have introduced $q_{r,i} = g^{\uparrow\downarrow}_{r,i} / 2\pi \hbar N_0 \Dperp$, where $\Dperp = 2\eF\tau/3\mperp$ is the out-of-plane diffusion constant, and $\tilde{q}_{1,2} = q_{1,2}\tanh(q_{1,2} \tR)$. An outline of the derivation is given in App.~\ref{App_SDE}. Equations (\ref{Eq_sy_abc})--(\ref{Eq_f_gamma}) explicitly describe how the spatially resolved spin polarization in an anisotropic Rashba metal depends on the magnetization direction of the adjacent ferromagnet. We wish to point out that these equations fully determine the magnetoresistance signals, as we shall see in the following section.

\section{Magnetoresistance}\label{Sec_Magnetoresistance}
The resistivity $\rho$ is defined by
\begin{equation}
E_x =\rho j_x \, ,
\end{equation}
where $E_x$ is the electric field, and $j_x = 1/\tR \int_0^{\tR} dz j_x(z) $ is the current density averaged over the thickness of the Rashba system. In the following, quantities without explicit $z$ dependence are considered as thickness-averaged.
Regarding the magnetoresistance, it is convenient to the split the resistivity,
\begin{equation}
\rho = \rho_0 + \Delta \rho(\bn) \, , \label{Eq_Def_Deltarho}
\end{equation}
where $\rho_0 \approx 1/\sigmaD$ is the resistivity for vanishing spin-mixing conductance, $\gr = \gi = 0$, and $\Delta \rho$ captures the magnetization dependence. 
From the generalized Boltzmann equation, see App.~\ref{App_Boltzmann}, one obtains
\begin{widetext}
\begin{equation}
 j_x(z) = \sigmaD E_x + 2e \left[ \frac{\lDP}{\taus} \left( 1- \frac{3\zeta}{4} \right) \theta_\mathrm{int}^\mathrm{sH} s^y(z) - \left( \theta_\mathrm{int}^\mathrm{sH} + \theta_\mathrm{ext}^\mathrm{sH} \right) j_y^z(z) + \frac{\mperp}{\mparallel} \theta_\mathrm{ext}^\mathrm{sH} j_z^y(z) \right]  \, . \label{Eq_jx}
\end{equation}
\end{widetext}
Loosely speaking, the first term in the square brackets corresponds to the spin galvanic or inverse Edelstein effect, the second term to the in-plane inverse spin Hall effect, and the third term to the out-of-plane inverse spin Hall effect. Interestingly, the relevant spin currents,
\begin{align}
j_y^z(z) &= \frac{\lDP}{\tauDP} s^y(z) +  \left(\theta_\mathrm{int}^\mathrm{sH} + \theta_\mathrm{ext}^\mathrm{sH} \right)\frac{\sigmaD E_x}{2e} \, , \\
j_z^y(z) &= - \Dperp \nabla_z s^y(z) - \theta_\mathrm{ext}^\mathrm{sH} \frac{\mparallel}{\mperp} \frac{\sigmaD E_x}{2e} \, ,
\end{align}
are completely determined by $\Delta s^y(z,\bn)$ regarding their dependence on the magnetization of the ferromagnet. Hence, the angular dependence of $j_x(z)$, and thus the magnetoresistance, can be traced back to $\Delta s^y(z,\bn)$. 

With the definition of the conductivity, $ j_x = \sigma E_x$, where $\sigma = \sigma_0 + \Delta \sigma(\bn)$, analogously to Eq.~(\ref{Eq_Def_Deltarho}), we obtain
\begin{equation}
\Delta \sigma(\bn) E_x = -2 e \Dparallel \qsg \Delta s^y(\bn) \, . \label{Eq_Deltasigma}
\end{equation}
Here, we have introduced 
\begin{equation}
\qsg = \frac{1}{\lDP} \Big[ \xi_\mathrm{int} \theta_\mathrm{int}^\mathrm{sH} + \xi_\mathrm{ext} \theta_\mathrm{ext}^\mathrm{sH} \Big] \, ,
\end{equation}
a wave number which represents the efficiency of the spin galvanic effect, with
\begin{equation}
\xi_\mathrm{ext} =1 - \qone \lDP \tanh (\qone \tR / 2) \, , \label{Eq_xi_ext}
\end{equation}
and $\xi_\mathrm{int}$ as defined in Eq.~(\ref{Eq_xi}). 
For a thin system, $\qone \tR \ll 1$, and assuming $\qone \lDP \lesssim 1$, the second term on the r.h.s.\ of Eq.~(\ref{Eq_xi_ext}) is negligible. Equivalently, the last term in the square brackets of Eq.\ (\ref{Eq_jx}), after averaging w.r.t.\ the thickness, is small, which means that the out-of-plane spin current $j_z^y$ does not contribute to the magnetoresistance, similar to a strictly 2D system.

Assuming $\sigma_0 \approx \sigmaD \gg \Delta \sigma$, the magnetization-dependent part of the resistivity is given by
\begin{equation}
\frac{\Delta \rho(\bn)}{\rho_0} \approx - \frac{\Delta \sigma(\bn)}{\sigmaD} \, . \label{Eq_Deltarho}
\end{equation}
We insert Eq.~(\ref{Eq_Deltasigma}) together with the thickness average of $\Delta s_{\alpha,\beta,\gamma}^y(z)$, Eq.~(\ref{Eq_sy_abc}), and obtain the magnetoresistance ratio 
\begin{equation} 
\frac{\Delta \rho_{\alpha,\beta,\gamma}}{\rho_0} =C f_{\alpha,\beta,\gamma}
\end{equation}
for the $\alpha$, $\beta$, and $\gamma$ scans, respectively. The magnitude of the effect is determined by
\begin{equation} 
C = \frac{\tanh(\qone \tR)}{ \qone\tR} \frac{\qsg}{\qisg} \left[ 1-  \tanh\left( \frac{\qone \tR}{2} \right) \theta_\mathrm{ext}^\mathrm{sH}  \frac{\qisg}{\qone} \right] \, ,
\end{equation}
with the ratio $\qsg/\qisg$ being quadratic in the spin Hall angles. However, due to the simultaneous contributions from $s^y$, $j_y^z$, and $j^y_z$, the magnetoresistance \emph{cannot} generally be expressed in terms of the square of a single total spin Hall angle $\theta_\mathrm{sH}$, as in the phenomenological approach.\cite{chen2013} Only in the special case where intrinsic spin-orbit coupling is negligible, the magnetoresistance can be expressed in terms of a single spin Hall angle squared.
In the following, we discuss the magnetoresistance for the representative limits of a purely damping-like torque, $\gi = 0 $, and a purely field-like torque, $\gr = 0$, respectively.

\subsection{Damping-like torque}
In the case of a vanishing imaginary part of the spin mixing conductance, $\gi=0$, which corresponds to a damping-like torque, the angular-dependent magnetoresistances are given by
\begin{align} 
\frac{\Delta \rho_\alpha}{\rho_0} &=  \frac{ C q_r  }{q_r + \tqone} \cos^2 \alpha \, , \label{Eq_rho_alpha_qr} \\
\frac{\Delta \rho_\beta}{\rho_0} &=  \frac{ C q_r   \cos^2 \beta}{q_r + \tqone - q_r\left(1 - \frac{\tqone}{\tqtwo}\right)\sin^2\beta }   \, ,  \label{Eq_rho_beta_qr} \\
\frac{\Delta \rho_\gamma}{\rho_0} &=  \frac{ C q_r }{q_r + \tqone}  \, . \label{Eq_rho_gamma_qr}
\end{align}
We see that $\Delta \rho_\gamma$ is constant, and that $\Delta \rho_\beta$ has a similar angular dependence as $\Delta \rho_\alpha$ for a wide range of parameters. More precisely, in the case $|1-\tqone/\tqtwo| \ll 1$, the $\sin^2\beta$ term in the denominator in Eq.~(\ref{Eq_rho_beta_qr}) leads to higher harmonics in $\beta$ of smaller magnitude,
\begin{equation}
\frac{\Delta \rho_\beta}{\rho_0} \approx    \frac{C q_r   }{q_r + \tqone} \left[ \cos^2\beta + \frac{ 1 -\tqone/\tqtwo }{4(1 +\tqone/q_r)}  \sin^2(2\beta) \right] \, .
\end{equation}
Apparently, the ratio $\tqone/\tqtwo$ determines the sign of the next-to-leading harmonic of the signal.

\begin{figure}[tb]
\centering
\includegraphics[width=\columnwidth]{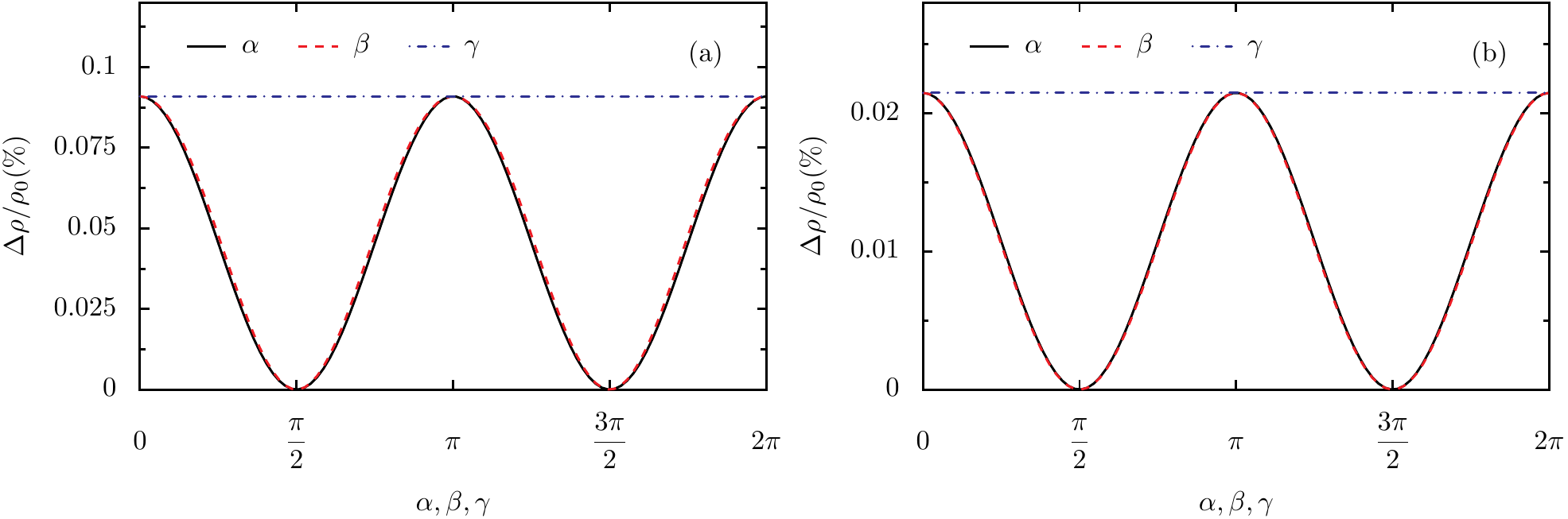}
\caption{Magnetoresistance for a damping-like torque as function of $\alpha$, $\beta$, and $\gamma$ with $\zeta=0.5$, $ q_r  \lDP= 0.5$, and $\tR= 2 \lDP$. The left panel, (a), corresponds to the case of strong Rashba spin-orbit coupling ($\tauDP/\taus = 0.5$, $\theta_\mathrm{int}^\mathrm{sH} = 0.1$, $\theta_\mathrm{ext}^\mathrm{sH} = 0.01$), and the right panel, (b), to the case of dominant extrinsic spin-orbit coupling ($\tauDP/\taus = 5$, $\theta_\mathrm{int}^\mathrm{sH} = 0.01$, $\theta_\mathrm{ext}^\mathrm{sH} = 0.1$). \label{Fig_rho_damping}}
\end{figure}

Figure \ref{Fig_rho_damping} shows the magnetoresistance according to Eqs.~(\ref{Eq_rho_alpha_qr})--(\ref{Eq_rho_gamma_qr}). 
Panel (a) corresponds to the case where Rashba spin-orbit coupling is large compared to the extrinsic  spin-orbit coupling, whereas (b) corresponds to the opposite limit. 
When Rashba spin-orbit coupling dominates the signal is larger by roughly an order of magnitude as compared to the extrinsic-dominated case. However, the angular dependence is very similar in the two regimes.

\subsection{Field-like torque}
In order to elucidate the effect of a purely field-like torque, we now neglect the real part of the spin mixing conductance, $\gr=0$. The angular-dependent magnetoresistances are then given by
\begin{align} 
\frac{\Delta \rho_\alpha}{\rho_0} &=   \frac{C  q_i^2  }{q_i^2+ \tqone \tqtwo} \cos^2 \alpha \, , \label{Eq_rho_alpha_qi} \\
\frac{\Delta \rho_\beta}{\rho_0} &=    \frac{C q_i^2   \cos^2 \beta}{ q_i^2 + \tqone^2 - q_i^2\left(1 - \frac{\tqone}{\tqtwo}\right)\sin^2\beta }   \, ,  \label{Eq_rho_beta_qi} \\
\frac{\Delta \rho_\gamma}{\rho_0} &=    \frac{C q_i^2 \left[ 1 -  \left(1 - \frac{\tqone}{\tqtwo}\right)  \sin^2 \gamma \right]}{ q_i^2 + \tqone^2 - q_i^2\left(1 - \frac{\tqone}{\tqtwo}\right)\sin^2\gamma } \, . \label{Eq_rho_gamma_qi}
\end{align}
The ratio $\tqone/\tqtwo$ defines the sign of the $\sin^2\gamma$ contribution in Eq.~(\ref{Eq_rho_gamma_qi}). It also determines whether the ratio of the amplitudes of $\Delta \rho_\alpha$ and $\Delta \rho_\beta$,
\begin{equation}
\frac{\Delta \rho_\alpha(0)}{\Delta \rho_\beta(0)} = 1- \frac{\tqone\tqtwo}{q_i^2 + \tqone\tqtwo} \left(1-\frac{\tqone}{\tqtwo} \right) \, , \label{Eq_ratio_amplitude}
\end{equation}
is larger or smaller than one. For $ \tqtwo \gg q_i$ and $ \tqone \gg q_i$, Eq.~(\ref{Eq_ratio_amplitude}) reduces to
\begin{equation}
\frac{\Delta \rho_\alpha(0)}{\Delta \rho_\beta(0)} \approx \frac{\tqone}{\tqtwo} \, , \label{Eq_ratio_amplitude_small_qi}
\end{equation}
such that the ratio $\tqone/\tqtwo$ can be read off directly from the measured amplitudes of the $\alpha$ and $\beta$ signals. Inserting the definitions of $\tqone$ and $\tqtwo$, Eqs.~(\ref{Eq_qperp}) and (\ref{Eq_qparallel}), into Eq.~(\ref{Eq_ratio_amplitude_small_qi}), we can solve for 
\begin{equation}
\frac{\tauDP}{\taus} = \frac{2\left( \frac{\Delta \rho_\alpha(0)}{\Delta \rho_\beta(0)} \right)^2 - 1}{1 - \zeta \left( \frac{\Delta \rho_\alpha(0)}{\Delta \rho_\beta(0)} \right)^2} \, , \label{Eq_ratio_extract_DP_s}
\end{equation}
or, in case $\tauDP \gg \taus$, directly extract the anisotropy parameter of the Elliott-Yafet spin relaxation,
\begin{equation}
\zeta \approx  \left(\frac{\Delta \rho_\alpha(0)}{\Delta \rho_\beta(0)} \right)^{-2} \, . \label{Eq_get_zeta}
\end{equation}
Analogous to the damping-like case, up to linear order in $(1-\tqone/\tqtwo)$, we can expand $\Delta \rho_\beta$ in terms of harmonics in $\beta$,
\begin{equation}
\frac{\Delta \rho_\beta}{\rho_0} \approx    \frac{C q_i^2  }{q_i^2 + \tqone^2} \left[ \cos^2\beta + \frac{1}{4} \left( \frac{  1 -\tqone/\tqtwo }{1 + \tqone^2 / q_i^2} \right) \sin^2(2\beta) \right] \, ,
\end{equation}
and similarly $\Delta \rho_\gamma$ can be expressed as
\begin{equation}
\frac{\Delta \rho_\gamma}{\rho_0} \approx   \frac{C q_i^2  }{q_i^2 + \tqone^2} \left[ 1  - \frac{\tqone^2}{q_i^2}  \left( \frac{  1 -\tqone/\tqtwo }{1 + \tqone^2 / q_i^2} \right) \sin^2\gamma \right] \, . \label{Eq_rho_gamma_qi_expand}
\end{equation}
We see that one can obtain the ratio $q_i/\tqone$ by dividing the amplitude of the second-harmonic of $\Delta \rho_\beta$ by the amplitude of the $\gamma$ scan of the magnetoresistance. 

\begin{figure}[tb]
\centering
\includegraphics[width=\columnwidth]{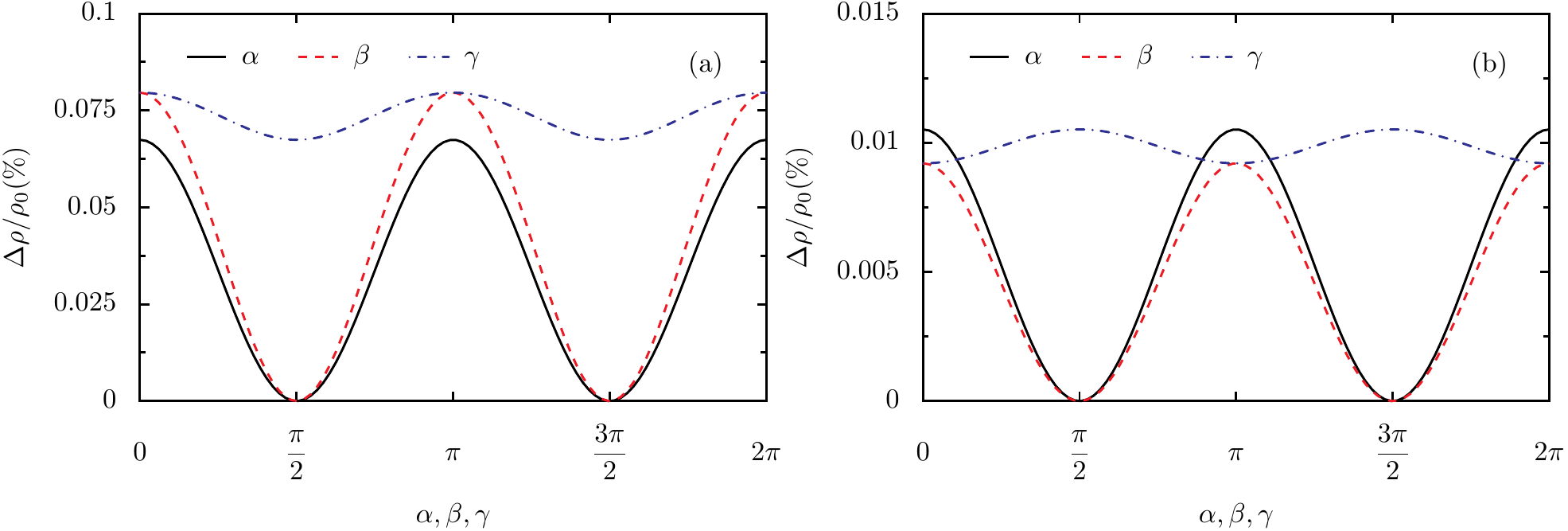}
\caption{Magnetoresistance for a field-like torque as function of $\alpha$, $\beta$, and $\gamma$ with $\zeta=0.5$, $ q_r  \lDP= 0.5$, and $\tR= 2 \lDP$. The left panel, (a), corresponds to the case of strong Rashba spin-orbit coupling ($\tauDP/\taus = 0.5$, $\theta_\mathrm{int}^\mathrm{sH} = 0.1$, $\theta_\mathrm{ext}^\mathrm{sH} = 0.01$), and the right panel, (b), to the case of dominant extrinsic spin-orbit coupling ($\tauDP/\taus = 5$, $\theta_\mathrm{int}^\mathrm{sH} = 0.01$, $\theta_\mathrm{ext}^\mathrm{sH} = 0.1$).  \label{Fig_rho_field}}
\end{figure} 

\subsection{Discussion}
First, we emphasize that our work allows a microscopic description of the magnetoresistance in anisotropic Rashba systems. The theory is not only applicable to real Rashba metals, but also to heavy-metal/ferromagnet bilayers, and substantially extends established phenomenological approaches.\cite{chen2013} The latter are contained in our results by setting $\zeta=1$, $\alpha_R = 0$, and $\theta_\mathrm{ext}^\mathrm{sH} \to \theta_\mathrm{sH}$. 

Second, we wish to stress the following two aspects: (1) the consideration of a mass anisotropy, and (2) the inclusion of Rashba spin-orbit coupling.
The mass anisotropy, point (1), leads to an anisotropic spin relaxation, even in the case of vanishing Rashba spin-orbit coupling $\alpha_R = 0$, and thus $\tqone/\tqtwo \neq 1$. In this case, according to Eq.~(\ref{Eq_f_gamma}) the $\gamma$ scan acquires a finite amplitude $\Delta \rho_\gamma$ when the imaginary part of the spin mixing conductance is nonzero, $\gi \neq 0$. Hence, using a ferromagnetic insulator, one can extract the ratio of the reduced spin relaxation lengths $1/\tqone$ and $1/\tqtwo$ by a precise measurement of $\Delta \rho_\gamma$. Indeed, experimental results for a Cu[Pt]/YIG bilayer structure, where the Cu/YIG interface is sputtered with Pt nanosize islands, show a noticeable oscillation in the $\gamma$ scan,\cite{zhou2018} which can be explained within our theory assuming a nonzero $\gi$ and $\tqone/\tqtwo > 1$, cf.\ Fig.~\ref{Fig_rho_field} (b). This effect is quite pronounced due to an enhancement of the anisotropy of the spin relaxation mechanism as the sputtered Pt exhibits Rashba spin-orbit coupling.\cite{zhou2018}
This directly brings us to point (2). There is evidence that thin Pt films also possess a strong Rashba spin-orbit coupling.\cite{ryu2016} In this case, the inverse spin galvanic effect strongly influences the spin transport, and the magnetoresistance signal cannot be interpreted as spin Hall magnetoresistance in the sense of a `simple' interplay between the spin Hall and the inverse spin Hall effect, resulting in $\Delta \rho \sim \theta_\mathrm{sH}^2$. Instead, one should focus on the spin polarization $s^y$, described by the wavenumbers $\qsg$ and $\qisg$, which represent the efficiency of the conversion of an electric field to a spin polarization and vice versa. Therefore, our theory is a generalization of previous approaches\cite{chen2013} which focus on the spin Hall angle and the spin currents.

Last but not least, our results compare favorably with experiments on hybrid structures consisting of spin-orbit active materials and a ferromagnetic metal. In these measurements, the $\gamma$ scan is usually explained by an additional contribution from the anisotropic magnetoresistance.\cite{avci2015, liu2015, kim2016} Note, however, that the measured signals also qualitatively agree with the magnetoresistance obtained in this work for a field-like torque, see Fig.~\ref{Fig_rho_field}. Since the spin mixing conductance is determined by interface properties, it is not obvious that its imaginary part is  always negligible. Therefore, special care is required when interpreting the measured signals. For example, the magnetoresistance in a Bi(15nm)/Ag/CoFeB trilayer, where a Rashba 2DEG is present at the Bi/Ag interface, shows a sign reversal in the oscillation of the $\gamma$ scan when comparing the low-temperature with the room-temperature measurements.\cite{nakayama2017} Qualitatively, the signals in the first case agree with Fig.~\ref{Fig_rho_field} (a), and in the second case with Fig.~\ref{Fig_rho_field} (b). Since $1/\tau$ is typically an increasing function of the temperature and $\tauDP/\taus \sim 1/\tau^2$, the ratio $\tqone/\tqtwo$ is also temperature dependent. Hence, Fig.~\ref{Fig_rho_field} (a) with $\tqone/\tqtwo < 1$ due to a small ratio $\tauDP/\taus = 0.5$ corresponds to the low-temperature regime and, vice versa, Fig.~\ref{Fig_rho_field} (b) with $\tqone/\tqtwo > 1$ to the high-temperature case.

\section{Conclusions} \label{Sec_conclusion}
We have presented a microscopic theory of the magnetoresistance in bilayer structures consisting of a Rashba metal and a ferromagnetic insulator, where the Rashba metal exhibits a mass anisotropy. Extrinsic spin-orbit coupling due to impurities has been taken into account via Elliott-Yafet spin relaxation, as well as side-jump and skew scattering. 
The mass anisotropy of the Rashba metal leads to an anisotropic Elliott-Yafet spin relaxation mechanism. Consequently, and enhanced by Dyakonov-Perel spin relaxation,  the spin diffusion equations contain two different spin relaxation lengths. 
Notably, the angular dependence of the magnetoresistance is fully determined by the current-induced spin polarization. In order to illustrate the relevance of the Rashba-metal/ferromagnet interface, we have considered a purely damping-like and a purely field-like torque, respectively. In both cases, the magnitude of the magnetoresistance is strongly enhanced when Rashba spin-orbit coupling is large compared to extrinsic contributions. Interestingly, for a field-like torque the $\gamma$ scan acquires a nonzero amplitude whose sign is determined by the ratio of the two spin relaxation lengths, and is thus directly related to the anisotropy of the spin relaxation. Due to the temperature dependence of the spin relaxation lengths, a sign change in the amplitude of the $\gamma$ scan is predicted which may explain the experimentally observed temperature dependence. A careful analysis of the experimental data will therefore provide important information concerning the anisotropy of the spin relaxation mechanism and its temperature dependence.

\begin{acknowledgments}
We acknowledge stimulating discussions with Christian Back, as well as financial support from the German Research Foundation (DFG) through TRR 80 and SFB 1277.
\end{acknowledgments}

\appendix

\section{Kinetic theory} \label{App_Boltzmann} 
We employ the generalized Boltzmann equation derived in Ref.~\onlinecite{gorini2010}. In the static case it reads
\begin{align} \label{Eq_Boltzmann}
 \frac{i}{\hbar} \frac{\bp}{\mparallel} \cdot \left[ \bcalA^a \frac{ \sigma^a}{2} , f \right] + \frac{p_z}{\mperp}  f +&\frac{1}{2} \left\{ \bcalF \cdot \nabla_\bp , f  \right\}  \nonumber \\
 &= I_0 + I_\mathrm{EY} + I_\mathrm{ext} \, .
\end{align}
Here, we have assumed that the system is homogeneous in the $xy$ plane and inhomogeneous for $z>0$ due to the attachment to the ferromagnet at $z=0$. The nonzero components of the SU(2) vector potential and the SU(2) Lorentz force with the electric field $E_x \bx$ are given by
\begin{align}
\mathcal{A}^x_y &= - \mathcal{A}^y_x = \frac{\hbar}{\lDP} \, , \\
\bcalF &= - e E_x \bx - \frac{\bp}{\mparallel} \times \bcalB^a \frac{\sigma^a}{2}  \, ,
\end{align}
where the only nonzero component of the SU(2) magnetic field $\mathcal{B}_i^a$ is $\mathcal{B}_z^z = - \hbar/\lDP^2$. The collision operators on the r.h.s.\ of the Boltzmann equation (\ref{Eq_Boltzmann}) describe momentum relaxation ($I_0$) with the relaxation rate $1/\tau$, Elliott-Yafet spin relaxation ($I_\mathrm{EY}$) associated with the relaxation rate $1/\taus$, and side-jump and skew scattering ($I_\mathrm{ext}$, see Ref.\ \onlinecite{raimondi2012} for details). More precisely, the Elliott-Yafet collision operator consists of
\begin{equation}
I^0_\mathrm{EY} = - \frac{1}{\taus} \bsigma \cdot (\Gamma \langle \bff \rangle) \, ,  \label{Eq_I_0}
\end{equation}
where $\langle \dots \rangle$ denotes the angular average and $\Gamma = \mathrm{diag}(1,1,\zeta)$ accounts for the anisotropy of Elliott-Yafet spin relaxation. In addition, the Elliott-Yafet collision operator yields the following linear in the SU(2) potential contributions:
\begin{widetext}
\begin{align}
I^\mathcal{A}_{\mathrm{EY}, s} = \frac{1}{N_0 \tau} \left(\frac{\lambda}{2\hbar}\right)^4  \mathcal{A}_i^a \varepsilon_{ijk} \varepsilon_{lmn}  \int \frac{d^3p'}{(2\pi\hbar)^3} & \delta(\epsilon_\bp - \epsilon_{\bp'}) \left(  f_{\bp'}^0  -  f_{\bp}^0 \right)  \nonumber \\
& \times \left[  p'_k p'_n p_m \delta_{jl} \sigma^a - p'_n  p_k p_m \left(\delta_{aj}\sigma^l + \delta_{al} \sigma^j - \delta_{jl}\sigma^a  \right) \right]  \, ,  \label{Eq_I_A_spin} \\
I^\mathcal{A}_{\mathrm{EY}, c} = \frac{1}{N_0 \tau} \left(\frac{\lambda}{2\hbar}\right)^4 \mathcal{A}_i^a \varepsilon_{ijk} \varepsilon_{lmn} \int \frac{d^3p'}{(2\pi\hbar)^3} & \delta(\epsilon_\bp - \epsilon_{\bp'}) p'_k p'_n p_m   \nonumber \\
& \times \left[ f_{\bp'}^b ( \delta_{al} \delta_{bj}  + \delta_{aj} \delta_{bl}) - \delta_{ab}\delta_{jl} \left(  f_{\bp'}^b  + f_{\bp}^b \right) \right]  \, ,  \label{Eq_I_A_charge}
\end{align}
where $\epsilon_\bp = (p_x^2 + p_y^2)/2\mparallel + p^2_z/\mperp$ is the band energy and $s$ ($c$) denotes a contribution to the spin (charge) sector. The above collision operators, Eqs.\ (\ref{Eq_I_0})--(\ref{Eq_I_A_charge}), are obtained by following the outline given in Ref.\ \onlinecite{gorini2010} with the self-energies
\begin{align}
\tilde{\Sigma}^0_\mathrm{EY} =& \frac{1}{2\pi \hbar N_0 \tau} \left( \frac{\lambda}{2\hbar} \right)^4 \int \frac{d^3 p'}{(2\pi\hbar)^3} \left[ \left( \bp \times \bp' \right)\cdot \bsigma \right] \tilde{G}(\bp') \left[ \left( \bp \times \bp' \right)\cdot \bsigma \right] \, , \label{Eq_App_Sigma_0} \\
\tilde{\Sigma}^\mathcal{A}_\mathrm{EY} =& \frac{1}{4\pi \hbar N_0 \tau}  \left( \frac{\lambda}{2\hbar} \right)^4 \big\lbrace \mathcal{A}_i^a \sigma^a , \nabla_{p_i}  \int \frac{d^3 p'}{(2\pi\hbar)^3} \left[ \left( \bp \times \bp' \right)\cdot \bsigma \right] \tilde{G}(\bp') \left[ \left( \bp \times \bp' \right)\cdot \bsigma \right] \big\rbrace \nonumber \\
&- \frac{1}{4\pi \hbar N_0 \tau} \left( \frac{\lambda}{2\hbar} \right)^4 \int \frac{d^3 p'}{(2\pi\hbar)^3} \left[ \left( \bp \times \bp' \right)\cdot \bsigma \right]  \big\lbrace \mathcal{A}_i^a \sigma^a , (\nabla_{p'_i}  \tilde{G}(\bp') ) \big\rbrace \left[ \left( \bp \times \bp' \right)\cdot \bsigma \right] \, , \label{Eq_App_Sigma_A}
\end{align}
\end{widetext}
where $\lbrace \cdot , \cdot \rbrace$ denotes the anti-commutator and $\tilde{G}$ is the locally covariant Green's function in Keldysh space. Although not explicitly indicated, the self-energies and Green's function are taken as impurity averaged, $\langle V(\br) V(\br') \rangle_\mathrm{imp} = (\hbar/2\pi N_0 \tau) \delta(\br - \br')$. 
For more details on the Elliott-Yafet collision operator we refer, for instance, to Refs.~\onlinecite{toelle2017,gorini2017}.

By performing the trace of the Boltzmann equation (\ref{Eq_Boltzmann}), multiplying with $p_x$, performing the momentum integration, and rearranging the terms one obtains the charge current $j_x$ as given in Eq.\ (\ref{Eq_jx}). In order to derive the spin diffusion equations (\ref{Eq_sx_Diff})--(\ref{Eq_sz_Diff}) we consider the trace of the Boltzmann equation after multiplication with the Pauli vector, $\mathrm{Tr}[\bsigma/2 \dots]$. From the resulting $3\times 3$ matrix equation we can obtain two equations for the spin density and the spin current: first, a direct integration over the momentum yields
\begin{equation} \label{Eq_continuity}
\Gamma \bs + \taus \nabla_z \bj_z - \taus \frac{\bcalA_i}{\hbar} \times \bj_i =  \by \theta^\mathrm{sH}_\mathrm{int} \frac{\tauDP}{\lDP} \left( 1-\frac{3\zeta}{4} \right)  \frac{\sigmaD E_x}{2e}   \, .
\end{equation}
Second, solving for $\bff$ in terms of $\langle  \bff \rangle$ and $f^0$, and performing the momentum integration after a multiplication by $p_i$ with $i=x,y,z$ yields the spin currents
\begin{align} 
\bj_x  &= - \frac{\lDP}{\tauDP} \by \times \bs  \, , \label{Eq_spin_currentx}  \\
\bj_y  &=  \frac{\lDP}{\tauDP} \bx \times \bs  + \bz \left( \theta^\mathrm{sH}_\mathrm{int} +  \theta^\mathrm{sH}_\mathrm{ext} \right) \frac{\sigmaD E_x}{2e}  \, , \label{Eq_spin_currenty}\\
\bj_z  &= - \Dperp \nabla_z \bs - \by \theta^\mathrm{sH}_\mathrm{ext} \frac{\mparallel}{\mperp} \frac{\sigmaD E_x}{2e} \, . \label{Eq_spin_currentz}
\end{align}
The spin diffusion equations (\ref{Eq_sx_Diff})--(\ref{Eq_sz_Diff}) now follow from inserting Eqs.\ (\ref{Eq_spin_currentx})--(\ref{Eq_spin_currentz}) into Eq.\ (\ref{Eq_continuity}).

\section{Spin diffusion equations} \label{App_SDE} 
In this appendix we briefly outline how to solve the spin diffusion equations (\ref{Eq_sx_Diff})--(\ref{Eq_sz_Diff}) for the current-induced spin polarization $s^y$ as described by Eqs.\ (\ref{Eq_sy_total})--(\ref{Eq_f_gamma}). Since we deal with decoupled differential equations, the general solution is easily obtained,
\begin{align}
s^x &=  a_1 e^{-\qone z} + a_2 e^{\qone z} \, , \label{Eq_sx_general} \\
s^y &=  b_1 e^{-\qone z} + b_2 e^{\qone z}  + s_0^y \, , \label{Eq_sy_general} \\
s^z &=  c_1 e^{-\qtwo z} + c_2 e^{\qtwo z}  \, , \label{Eq_sz_general}
\end{align}
respectively. The boundary conditions as given in the main text, $\bj_z(\tR)=0$ and Eq.\ (\ref{Eq_STT}), can be applied to Eqs.~(\ref{Eq_sx_general})--(\ref{Eq_sz_general}) by employing Eq.\ (\ref{Eq_spin_currentz}). Considering first $\bj_z(\tR)=0$, we can reduce the number of unknown parameters,
\begin{align}
s^x &=  a \cosh\boldsymbol{(} \qone(\tR - z) \boldsymbol{)} \, , \label{3D_Eq_sx_a} \\
s^y &=  b \cosh\boldsymbol{(} \qone(\tR - z) \boldsymbol{)}  +  \frac{ \theta^\mathrm{sH}_\mathrm{ext} \sigmaD E_x}{2e\qone\Dparallel} \sinh\boldsymbol{(} \qone(\tR - z) \boldsymbol{)} + s_0^y \, , \label{3D_Eq_sy_b} \\
s^z &=  c \cosh\boldsymbol{(} \qtwo(\tR - z) \boldsymbol{)}  \, . \label{3D_Eq_sz_c}
\end{align}
It is now convenient to consider first the case without ferromagnet, i.e., $\bj_z(0)=0$. In this case, the spin density $s^y$ is given by
\begin{equation}
s^y(z)\big|_{q_r = q_i = 0} = s_0^y + \Delta s^y_{sc}(z) \, ,
\end{equation}
with $ \Delta s^y_{sc}$ as given in Eq.~(\ref{Eq_ssc}). In the presence of the ferromagnet, the spin density can be written as follows:
\begin{equation} \label{Eq_sy_almost_done}
\bs(z,\bn) = \left[ s^y_0 + \Delta s_\mathrm{sc}^y(z) \right] \by + \Delta \bs(z,\bn) \, ,
\end{equation}
where
\begin{equation}
\Delta \bs(z,\bn)  = 
\begin{pmatrix}
\tilde{a} \cosh\boldsymbol{(} \qone(\tR - z) \boldsymbol{)} \\
\tilde{b} \cosh\boldsymbol{(} \qone(\tR - z) \boldsymbol{)} \\
\tilde{c} \cosh\boldsymbol{(} \qtwo(\tR - z) \boldsymbol{)}
\end{pmatrix} \, . \label{Eq_Deltabs_general}
\end{equation}
By splitting the spin current $\bj_z$ similar to Eq.~(\ref{Eq_sy_almost_done}), the application of the boundary condition (\ref{Eq_STT}) leads to
\begin{align}
\begin{pmatrix}
\qone \tilde{a} \sinh(\qone \tR) \\
\qone \tilde{b} \sinh(\qone \tR) \\
\qone \tilde{c} \sinh(\qtwo \tR) 
\end{pmatrix}
=&\Dperp q_r \bn \times \big[ \bn \times \bs(z=0) \big]  \nonumber \\
&+ \Dperp q_i \bn \times \bs(z=0)  \label{App_3D_Eq_getb}
\end{align}  
with
\begin{equation}
\bs(z=0) = \left[ s^y_0 + \Delta s_\mathrm{sc}^y(0) \right] \by +
\begin{pmatrix}
 \tilde{a} \cosh(\qone \tR) \\
 \tilde{b} \cosh(\qone \tR) \\
 \tilde{c} \cosh(\qtwo \tR) 
\end{pmatrix} \, .
\end{equation}
The remaining task is now to solve for $\tilde{b}$. One way is to parametrize $\bn$ in spherical coordinates and multiply Eq.\ (\ref{App_3D_Eq_getb}) by the rotation matrix $\mathbb{D}$ with the property $\mathbb{D} \bn = \bx$. By a proper choice of the spherical angles regarding the $\alpha$, $\beta$, and $\gamma$ scans, respectively, the solution for $\tilde{b}$ of the resulting system of linear equations finally yields the spin density $\Delta s^y_{\alpha,\beta,\gamma}$ as given by Eqs.~(\ref{Eq_sy_abc})--(\ref{Eq_f_gamma}).

\bibliography{3D-Rashba}

\end{document}